\documentclass[conference]{IEEEtran}

\usepackage{fancyvrb}
\usepackage{amsmath}

\newcommand{\q}[1]{\lq\lq{}{}#1\rq\rq{}{}}

\begin{document}

\title{
    Towards a Ranking Model for Semantic Layers\\over Digital Archives
}

\author{
    \IEEEauthorblockN{
        Pavlos Fafalios,
        Vaibhav Kasturia,
        Wolfgang Nejdl}
    \IEEEauthorblockA{
        L3S Research Center, Hannover, Germany\\
        \{fafalios, kasturia, nejdl\}@l3s.de}
}

\maketitle

\begin{abstract}
Archived collections of documents (like newspaper archives)
serve as important information sources for historians, journalists, sociologists
and other interested parties.
Semantic Layers over such digital archives allow describing and publishing metadata and semantic
information about the archived documents in a standard format (RDF),
which in turn can be queried through a structured query language (e.g., SPARQL).
This enables to run advanced queries by combining
metadata of the documents (like {\em publication date}) and content-based semantic information
(like {\em entities} mentioned in the documents).
However, the results returned by structured queries
can be numerous and also they all equally match the query.
Thus, there is the need to rank these results in order to promote
the most important ones.
In this paper, we focus on this problem
and propose a ranking model that considers and combines:
i) the relativeness of documents to entities,
ii) the timeliness of documents, and
iii) the relations among the entities.
\end{abstract}

\section{Introduction}
Despite the increasing number of digital archives worldwide
(like newspaper and web archives), the absence of efficient and meaningful
exploration methods still remains a major obstacle in the way of
turning them into a usable source of information \cite{calhoun2014exploring}.
Semantic models
try to solve this problem by offering a vocabulary for describing
and publishing in the standard RDF format, metadata (e.g.,
publication date) and semantic (e.g., mentioned
entities) information  about a collection of archived documents. The produced
{\em Semantic Layers} allow running advanced {\em entity-centric}
queries requesting complex information related to some entities,
concepts or events and to some specific metadata values
\cite{fafalios2017SemLayer}. As an example, we can access a Semantic
Layer over a newspaper archive and find articles of a specific time
period discussing about a specific category of entities (e.g., {\em
philanthropists}) or about entities sharing some characteristics
(e.g., {\em lawyers born in Germany}). Such advanced information
needs can be directly expressed through SPARQL queries (unfriendly
for end-users) or through a user-friendly interactive interface
which transparently transforms user interactions to SPARQL queries
(e.g., a faceted browsing interface \cite{tzitzikas2016faceted}).
However, the results returned by such queries can be numerous and
moreover they all equally match the query. Thus, there arises the need
to rank them for discovering and returning to the user the most
important ones. An effective ranking method should consider the
different factors that affect the importance of a document to the
information need, relying at the same time only on the data
available in the semantic layer (i.e., without accessing documents' full
contents).

In this paper, we focus on this problem and
propose a model for ranking archived documents
returned by a structured query over a semantic layer.
The proposed model jointly considers the following aspects:
i) the {\em relativeness} of a document with respect to the entities of interest,
ii) the {\em timeliness} of document's publication date,
iii) the temporal {\em relatedness} of the entities of interest with other entities
mentioned in the document.
The idea is to promote documents that
mention the entities of interest many times,
that have been published in important (for the entities of interest) time periods, and that
mention many other entities co-occurring frequently
with the entities of interest in important time periods.
For example, in case we want to rank articles of 1990
discussing about {\em Nelson Mandela},
we want to favor articles that
i) mention {\em Nelson Mandela} multiple times in their text,
ii) have been published in important time periods for {\em Nelson Mandela}
(e.g., February 1990 since during that period he was released from prison), and
iii) mention other entities that seem to be important for {\em Nelson Mandela}
during important time periods
(e.g., {\em Frederik Willem de Klerk} who was
South Africa's State President in February 1990).

\section{Ranking Model}

\subsection{Problem Definition}

In our problem, an {\em entity} is anything with a distinct and meaningful existence
that also has an \q{identity} expressed through a unique ID (e.g., a Wikipedia URI).
This does not only include persons, locations, etc., but also
concepts (e.g., {\em democracy}) and
events (e.g., {\em 2010 Haiti earthquake}).

Given a collection of archived documents $D$,
a set of entities $E_D$ mentioned in documents of $D$,
and a SPARQL query $Q$ requesting documents from $D$
published within a set of {\em time periods} $P_Q$
of a fixed granularity $\Delta$ (e.g., day or week)
and related to one or more {\em Entities of Interest (EoI)} $E_Q \subseteq E_D$
with logical {\tt AND} (mentioning all EoI) or {\tt OR} (mentioning at least one EoI) semantics,
the problem is how to rank the documents $D_Q \subseteq D$ that (equally) match $Q$.

Figures \ref{fig:modelingExampleQ1} and \ref{fig:modelingExampleQ2}
show examples of SPARQL queries
requesting documents from a semantic layer over a newspaper archive
(see \cite{fafalios2017SemLayer} for more information about
the semantic layer).
The query in Fig. \ref{fig:modelingExampleQ1} requests articles
published in 1990 and
discussing about the entities {\em Nelson Mandela}
and {\em Frederik Willem de Klerk}
({\tt AND} semantics),
while the query in Fig. \ref{fig:modelingExampleQ2} requests
articles of 1990 mentioning {\em state presidents of South Africa}
({\tt OR} semantics).
Our objective is to rank the documents returned by
such SPARQL queries.

\begin{figure}[th]
\centering \scriptsize
\begin{Verbatim}[frame=lines,numbers=left,numbersep=1pt]
SELECT DISTINCT ?article WHERE {
 ?article dc:date ?date FILTER(year(?date) = 1990) .
 ?article schema:mentions ?entity1, ?entity2 .
 ?entity1 oae:hasMatchedURI dbr:Nelson_Mandela .
 ?entity2 oae:hasMatchedURI dbr:F._W._de_Klerk }
\end{Verbatim}
\vspace{-4mm}
\caption{Query requesting articles of 1990 mentioning
{\em Nelson Mandela} and {\em Frederik Willem de Klerk} ({\tt AND} semantics).}
\label{fig:modelingExampleQ1}
\vspace{3mm}
\centering \scriptsize
\begin{Verbatim}[frame=lines,numbers=left,numbersep=1pt]
SELECT DISTINCT ?article WHERE {
 ?article dc:date ?date FILTER(year(?date) = 1990) .
 ?article schema:mentions ?entity .
 ?entity oae:hasMatchedURI ?entURI .
 ?entURI dc:subject dbc:State_Presidents_of_South_Africa }
\end{Verbatim}
\vspace{-4mm}
\caption{Query requesting articles of 1990 discussing
about {\em state presidents of South Africa} ({\tt OR} semantics).}
\label{fig:modelingExampleQ2}
\end{figure}

\subsection{Modeling}

\noindent
{\bf Relativeness.}
We consider that if the EoI are mentioned
in a document many times, the document should receive a high score
since its topic may be about these entities.
The term frequency (in our case {\em entity frequency}) is a classic numerical statistic
that is intended to reflect how important
a word is to a document \cite{leskovec2014mining}.

For the case of {\tt AND} semantics (\q{$\wedge$}),
the {\em relativeness} score of a document $d \in D_Q$ can be
simply defined as:
\begin{equation}
\small
ScoreD_{\wedge}(d) = \frac{\sum_{e \in E_Q}{count(e, d)}}{\sum_{e' \in E_d}{count(e', d)}}
\end{equation}
where $E_d \subseteq E_D$ is the set of entities mentioned in $d$ and
$count(e, d)$ is the number of occurrences of an entity $e$ in $d$.

For the case of {\tt OR} semantics (\q{$\vee$}),
we can also consider the number of different EoI mentioned in the document
(since a document does not probably contain all the EoI as in the case of {\tt AND} semantics).
In that case, the {\em relativeness} score of a document $d \in D_Q$ can be defined as follows:
\begin{equation}
\small
ScoreD_{\vee}(d) = \frac{\sum_{e \in E_Q}{count(e, d)}}{\sum_{e' \in E_d}{count(e', d)}} \cdot \frac{|E_d \cap E_Q|}{|E_Q|}
\end{equation}
This formula favors documents mentioning multiple times many of the EoI.

\vspace{2mm} \noindent
{\bf Timeliness.}
A time period of granularity $\Delta$ can be considered important
for the EoI, if there is a relatively large number of documents
mentioning the EoI during that period.
For a time period $p \in P_Q$, we consider the following
{\em timeliness} score:
\begin{equation}
ScoreP(p) = \frac{|D_{p} \cap D_Q|}{|D_Q|}
\end{equation}
where $D_{p} \subseteq D$ is the set of documents published during $p$.

\vspace{2mm} \noindent
{\bf Relatedness.}
Entities that are co-mentioned frequently with the EoI
in important time periods are probably important for the EoI.
For example, {\em Apartheid} was an important concept
related to {\em Nelson Mandela} during 1990.
Thus, articles discussing for both {\em Apartheid} and {\em Nelson Mandela}
should be promoted.
However, there may be also some general entities (e.g., {\em South Africa} in our example)
that co-occur with the EoI in almost all documents (independently of the time period).
Thus, we should also avoid over-emphasizing documents mentioning such \q{common} entities.
First, we consider the following {\em relatedness} score
of an entity $e \in E_D \setminus E_Q$:

\begin{equation}
\begin{split}
\small
ScoreE(e) = & ~ idf(e) \cdot \sum_{p \in P_Q}{(ScoreP(p) \cdot \frac{|D_{e, p} \cap D_Q|}{|D_p \cap D_Q|})}\\
= & ~ idf(e) \cdot \sum_{p \in P_Q}{\frac{|D_{e, p} \cap D_Q|}{|D_Q|}}
\end{split}
\end{equation}
where $D_{e, p} \subseteq D$ is the set of documents mentioning $e$ and published in the time period $p$,
and $idf(e)$ is the inverse document frequency of $e$, defined as:
\begin{equation}
\label{frml:idf}
\small
idf(e) = 1 - \frac{|D_e \cap (\cup_{e' \in E_Q}{D_{e'}})|}{|\cup_{e' \in E_Q}{D_{e'}}|}
\end{equation}
where $D_e \subseteq D$ denotes the set of documents mentioning $e$.

This {\em relatedness} formula considers the percentage of
documents in which the entity
co-occurs with the EoI in important time periods.

\vspace{2mm} \noindent
{\bf Joining the Models.}
We can now join the above models and derive a final score
for a returned document $d \in D_Q$:
\begin{equation}
\small
S(d) = ScoreP(p_d) \cdot ScoreD(d) + \beta \frac{\sum_{e \in E_d \setminus E_Q}{ScoreE(e)}}{|E_d|}
\end{equation}
where $p_d \in P_Q$ is the time period in which $d$ was published,
and $\beta$ is a decay factor for controlling the effect of relatedness.

\section{Conclusion}
\label{sec:concl}

We have introduced a model for ranking
documents returned by querying
a semantic layer over an entity-annotated archived collection of documents.
An important characteristic of our approach is that
it only exploits the data of the semantic layer (i.e., its RDF triples)
and thereby it can be directly applied even at query-execution time.
In future, we will extensively evaluate the proposed model
and the effect of each of its components.
We also plan to investigate how this model
can be applied in web archives, where the
publication date is not usually available.

\vspace{2mm} \noindent
{\bf Acknowledgements}.
The work was partially funded by the
European Commission for the ERC Advanced Grant ALEXANDRIA (No. 339233).

\bibliographystyle{IEEEtran}
\bibliography{BIB}

\end{document}